\documentclass{elsart}
\usepackage[dvips]{graphicx}
\usepackage{color}
\usepackage{setspace}

\begin{document}
\begin{frontmatter}
\title{Atomic and electronic structure of ultra-thin Al/AlO$_{\textrm{\bf x}}$/Al interfaces}
\author{M. Die\v{s}kov\'{a}},
\author{M. Kon\^{o}pka}, and 
\author{P. Bokes\corauthref{cor}}
\ead{Peter.Bokes@stuba.sk}
\corauth[cor]{Corresponding Author.}
\address{Department of Physics, Slovak University of Technology FEI STU,
        Ilkovi\v{c}ova 3, 812 19 Bratislava, Slovak Republic}
%\author[FEI]{M. Die\v{s}kov\'{a}},
%\author[FEI]{M. Kon\^{o}pka}, and 
%\author[FEI]{P. Bokes\corauthref{cor}}
%\ead{Peter.Bokes@stuba.sk}
%%
%\corauth[cor]{Corresponding Author.}
%\address[FEI]{Department of Physics, Slovak University of Technology FEI STU,
%        Ilkovi\v{c}ova 3, 812 19 Bratislava, Slovak Republic}
%
%\maketitle
\begin{abstract}
Interfaces between metals based on AlO$_{x}$ represent the 
most popular basis for Josephson junctions or, more recently, 
also for junctions exhibiting substantial tunneling magneto-resistance. 
We have performed a computational study of possible local geometric structures 
of such interfaces at the ab-initio DFT/GGA level of approximation 
to complement recent experimental data on ultra-thin AlO$_{x}$-based interfaces. 
We present two competing structures that we characterize with their electronic 
properties: fragmentation and interface energies. 
\end{abstract}
\end{frontmatter}
\section{Introduction}
Improvement in performance of Josephson junctions or magnetic tunnel 
junctions based on aluminum oxide tunneling barrier critically 
depends on the quality of the interface Al/AlO$_x$/Al. It is desirable to 
produce highly transparent, i.e. ultra-thin interfaces ($< 1$nm) to achieve 
large critical currents while keeping the structures with sufficiently low
density of interfacial defects. Several recent attempts to characterize
such an ultra-thin barrier has been recently published. Rippard 
{\it et al.}~\cite{Rippard02} studied the dependence of the 
transmission spectrum of a ultra-thin interface ($0.6 - 1.5$nm) on the
oxidation, finding a conduction band within the oxide only 
$1.2$eV above the Fermi energy (effective barrier), which has been
argued to origin from a disorder in the barrier. Regions of 
similar barriers, referred to as ``hot spots'', were also observed by
Gloos {\it et al.}~\cite{Gloos03}. An interesting study by Tan {\it et al.}~\cite{Tan05} 
points to a novel stabilisation in the oxidation process and 
an inherent instability in AlO$_x$ films. They find that two structures 
- one corresponding to $x\approx1.2$ and the other with $x\approx1.0$ - can 
be switched using an electron gun, by exposure to oxygen or by 
overlaying the oxide with a metal having the work-function different from Al. 
Even though most of these barriers are known to be disordered, 
successful growth of an epitaxial AlO$_x$ barriers, which is an ultimate goal, 
bas been also reported~\cite{Mizuguchi05}.

Theoretical studies, closest in nature to the Al/AlO$_x$/Al interface, 
are concerned with oxidisation of aluminum surface, an area studied 
primarily due to its importance in corrosion. Particularly relevant 
for our investigation are studies of the oxidised $(111)$ surface 
of aluminum by Jennison {\it et al.}~\cite{Jennison99,Jennison00}. 
In their work the most energetically stable surface was found to 
be composed of an almost 2D layer of oxide with composition of 
Al$_2$O$_3$, separated from the bulk aluminum by a layer of chemisorbed 
oxygen. The resulting width of the oxide they found, $d\sim0.5$nm, 
approximately coincides with the estimated widths of the 
thinnest Al/AlO$_x$/Al interfaces found by Gloos~\cite{Gloos03}. 
This similarity suggests to take the 1x1 structure of the $(111)$ 
surface considered by Jennison as a starting candidate to represent 
the simplest ultra-thin interface. 

\section{Atomic structures}
\begin{figure}[p]
{\par\centering \includegraphics[width=6cm]{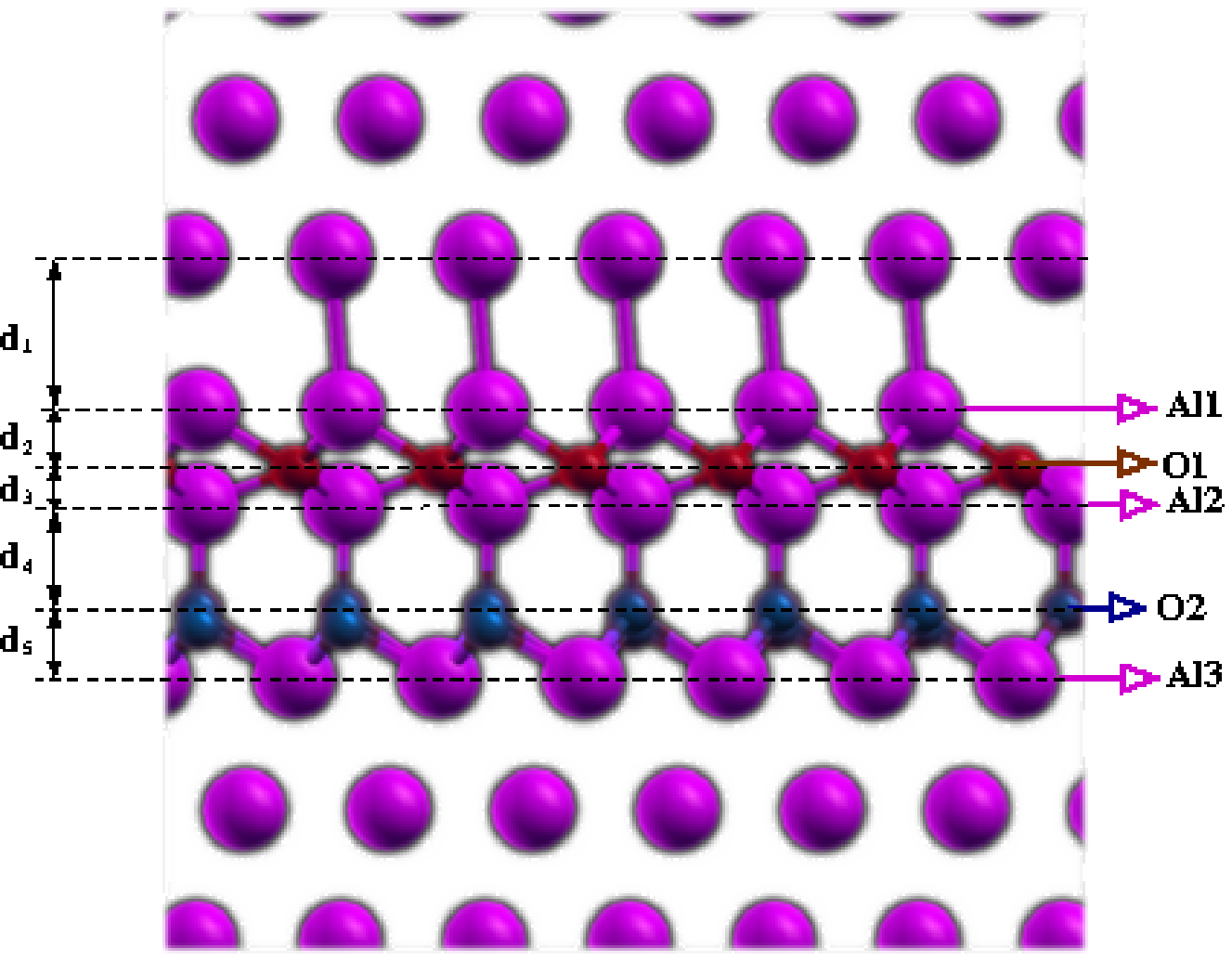}\par}
\caption{The asymmetric structure. O$2$ are the chemisorbed oxygen
atoms bound to a full (111) layer of bulk aluminum Al$3$. 
1/3 filling of Al$1$, fully filled layer of O$1$ and 2/3 filling
of Al$2$ layer represent the actual oxide.}
\label{fig1}
\end{figure}

Two different geometries of the interface were considered, both based on the 
1x1 surface supercell of an ideal (111) surface of Al. The first one
was build starting from a geometry of the ultra-thin oxidised surface
described by Jennison {\it et al.}~\cite{Jennison99,Jennison00}. By adding 
more atoms of aluminum and subsequent geometry optimisation combined with 
molecular dynamics we have identified the {\it asymmetric structure} A 
(see Fig.~\ref{fig1}). We have checked that the structure was in a true energy
minimum also for a 2x1 supercell. The interface consists of a chemisorbed
oxygen (atoms O$2$ in Fig.~(\ref{fig1})) bound to a full (111) layer 
of Al surface. Above the chemisorbed oxygen we find a 2/3-filled Al layer (Al$2$ atoms, 
hexagonal structure without the centers) forming with a full oxygen layer and 1/3-filled 
Al layer (Al$1$ atoms) an oxide AlO$_x$ with stoichiometry $x=1.0$. The geometrical 
parameters are given in the table~(\ref{tab-1}). The width of the interface, which we 
define as the vertical distance between the first non-bulk layer (Al$1$) and the 
last bulk-like layer (Al$3$) is $d^{tot}_A=4.53$\AA.

\begin{figure}[p]
{\par\centering \includegraphics[width=6cm]{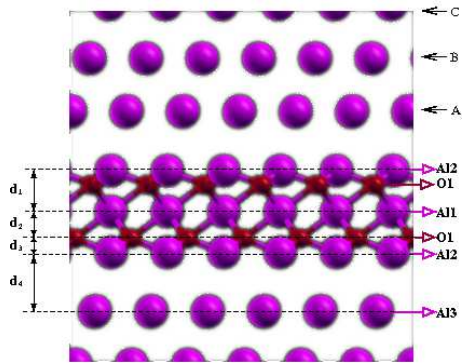}\par}
\caption{Symmetric structure $S$. The structure is symmetrically 
ordered with respect to the 1/3 filled plane of aluminum  atoms Al$2$. The letters
A, B, and C indicate the stacking of the (111) planes in the electrode.}
\label{fig2}
\end{figure}

The above mentioned optimisations with several Al ad-atoms indicated 
that there is a tendency for the Al$3$ atoms to move up, towards 
the oxidised surface. To help the optimisation to explore the possibilities 
we have utilised a well known ductile character of gold as well as its almost 
identical bulk lattice constant - we have exchanged the overlayer 
of Al with Au. However, to form a flat (111) surface we had to add 
two more Au atoms. The resulting structure was optimised into the final 
{\it symmetric structure} (S) shown in the Fig.~(\ref{fig2}). In contrast 
with structure A, the chemisorbed oxygen layer is missing and we find 
the AlO$_x$ oxide present in the upper half of the asymmetric interface instead, 
sharing the 1/3-filled Al layer with its mirror image. This can be directly checked 
by comparing the bond lengths and geometrical parameters given in the Table~(\ref{tab-1}).
Due to this sharing of the 1/3-filled layer the overall stoichiometry of S is $x=6/5=1.2$
The width of the interface S is found to be $d^{tot}_S=6.12$\AA.

In both structures there are also other geometric alternatives due to the miscellaneous 
possibilities of stacking of bulk alumina atoms. There are three different layers - A, B, C -
of bulk $(1 1 1)$ aluminum and by swaping their order one can obtain new geometries. 
Examples we considered are $CBA/BAC$ (see Fig.~(\ref{fig2})) which is the S-structure, 
$ABC/CAB$ and $CAB/BAC$. 

Our calculations were performed using ABINIT~\cite{abinit} and 
CASTEP~\cite{castep} programs for total energies per unit 
interface and fragmentation energies respectively. Within ABINIT 
we used Fritz-Haber-Institute pseudopotentials, which provide 
good convergence with plane wave cutoff of 60 Ry. The Brillouin 
zone integration was performed using 5 irreducible k-points 
within the Monkhorst-Pack (MP) mesh. Within CASTEP (calculation
of fragmentation energies), Vanderbilt ultra-soft pseudopotential 
with plane-wave cutoff 25 Ry and 2x2x1 MP mesh. The exchange 
and correlation were described at the level of general gradient 
approximation (GGA). Geometries were optimised until the RMS forces 
were $<0.05$eV/\AA.

\section{Interface Energy}
The ground state structure can be characterised with a minimal interface energy 
$\Delta E$. The latter is defined as the difference between the total energy 
per supercell of a concerned and some reference system. If one of these two 
systems has $N$ more metalic atoms inside the interface, as it is the case here, 
we need to compensate for this difference by adding an appropriate bulk energy 
of these $N$ atoms. This is motivated by the fact that bulk represents a reservoir 
that supplies atoms into the interface while it is being transformed between 
alternative interface geometries/stoichiometries.

Since the $S$ structure has two additional aluminum atoms\footnote{In fact, there 
is one Al atom less inside the S interface compared to structure A. However, the 
structure S in our calculation has an extral full (111) bulk plane of Al which 
contains 3 atoms per surface cell, hence we have excess of two atoms in total.}, 
we calculate the interface energy as
\begin{equation} 
	\Delta E = E_{S} - ( E_{A}  + 2 E_{Al} ),
\end{equation}
where $E_{Al}$ is the energy per atom in an ideal bulk aluminum, calculated 
with identical numerical precision as the interface supercell. The error 
introduced by a slight difference in the numerics has been checked to 
be $\leq 0.005$eV.

The resulting interface energy is $\Delta E = 0.072$eV$>0$, i.e.
the asymmetric structure is marginally lower in energy and hence 
it is more likely to be realised. Practically this means that in 
typically prepared disordered interfaces the local geometric and 
electronic structure of the ``asymmetric'' type will be more frequent 
than the one of the ``symmetric type''. Furthermore, due to their 
energetic proximity, it is not surprising that one can switch 
between them, as it has been indeed experimentally done by 
Tan {\it et al.}~\cite{Tan05} by means of electron bombardment or by 
overcoating with metals having different work function (Au in our case 
and Y, Nb, and Co by Tan {\it et al.}~\cite{Tan05}). The calculated 
interface energy as well as the fact that using the Au electrode
in place of the Al one drives the system towards the transition 
$A \rightarrow S$, represents further evidence that the experimentally 
observed behaviour by Tan {\it et al.} corresponds to the transition 
between local asymmetric and symmetric structures presented in our paper.

We have further examined the role of stacking of the fcc (111) planes in 
the metal on the interface energy. If the above treated structure S,
with $CBA/BAC$ stacking, represents the reference value, we obtain
$E_{ABC/CAB}  - E_{S} = 0.10$eV and $E_{CAB/BAC} - E_{S} = 0.28$eV.
Therefore, the different order of (111) planes of one electrode 
with respect to the structure S leads to noticeable differences 
in the interface energy. However, all the candidates we checked 
were above the energy of the symmetric interface.

\section{Fragmentation energies}
A useful criterion for comparison of the stability of different oxide structures 
represent fragmentation energies. In our particular case both examined 
structures have the same area of the surface cell which simplifies the evaluation.
We define fragmentation energy of a slab as
$$
E_{\rm frag} = E_{\rm lower} + E_{\rm upper} - E_{\rm whole} \ ,
$$
where $E_{\rm whole}$ is the energy of the whole, fully optimized structure 
(slab) in the supercell and $E_{\rm lower}$ and $E_{\rm upper}$ are similarly
obtained energies of the two parts of the whole structure (fully optimized 
to the nearest local minimum and in the same supercell). The fragmentation 
energy depends on the location of the cut. In this way we can obtain many 
different fragmentation paths, each one with its own energy. We are interested 
in the lowest-energy pathway which limits the stability of a given structure. 

Assume we have sereval different structures, $s_1, s_2, ..., s_N$, each one
having its weakest path with energy $E_s^{\rm min}$. Based on the fragmentation 
energies we define the most stable structure as the one with the 
highest $E_s^{\rm min}$.

\begin{figure}[p]
\begin{minipage}[t]{0.99\textwidth}
\centering
\begin{minipage}{0.45\textwidth}
\includegraphics[height=0.3\textheight]{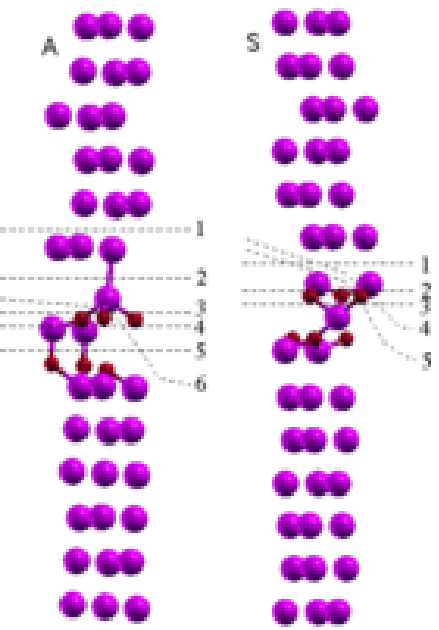}
\end{minipage}
\begin{minipage}{0.45\textwidth}
\begin{tabular}{|c|c|c|}
\hline
path & A    & S    \\ \hline \hline
  1  & 2.48 & 1.27 \\ \hline
  2  & 1.92 & 7.52 \\ \hline
  3  & 2.85 & 2.31 \\ \hline
  4  & 3.70 & 2.44 \\ \hline
  5  & 2.24 & 10.4 \\ \hline
  6  & 3.76 & ---  \\ \hline
\end{tabular}
\end{minipage}
\end{minipage}
\caption{Left panel: Images of $A$ and $S$ structures with fragmentation cuts 
shown schematically. Right panel: Calculated fragmentation energies in eV.}
\label{fig3}
\end{figure}

We have examined a limited set of fragmentation paths for the two
structures A and S, indicated in the Fig.~(\ref{fig3}). Corresponding 
fragmentation energies are listed in the table inside the figure. 
Thus, according to the given criterion and the limited set of examined 
fragmentation cuts, we conclude that the structure A is more stable.

\section{Conclusions}
In conclusions, we have presented optimised geometries 
for the Al/AlO$_{x}$/Al interface. Two geometries - one asymmetric with 
chemisorbed oxygen (A) and one with symmetrical arrangement (S) - 
were found to differ only by $E_{S}-E_{A} \approx 0.07$eV 
(per interface cell), yet they are principally 
different in stoichiometry ($x=1.0$ and $x=1.2$ respectively) 
and interface width ($d=4.53$\AA and $d=6.12$\AA respectively). 
The asymmetric structure is not only energetically more favorable,
but it is also more stable against fragmentation when compared to the symmetric 
structure. These findings are in agreement with recent experiments~\cite{Tan05}, 
identifying reproducible switching between two different oxide films, but indicate 
that in contrast to their interpretation, the morphological change
takes place deeper in the interface between the oxide and substrate. 
Second, in agreement with this experiment, we have confirmed that 
the two geometries can be switched by depositing metal with work function 
different from the one of the substrate Al.

\begin{ack}
The authors acknowledge the Center for Computational Materials Science
at the Slovak University of Technology for computational resources.
This research was supported by the Slovak grant agency VEGA 
(project No. 1/2020/05) and the NATO Security Through Science 
Programme (EAP.RIG.981521).
\end{ack}

%a table
\begin{table}[p]
\centering
\begin{tabular}{ccccccccc}
  & d$_1$ & d$_2$ & d$_3$ & d$_4$ & d$_5$ & d(O$2$-Al$3$) & d(O$1$-Al$2$)& d(O$1$-Al$1$) \\ \hline
A & 2.6   & 1.0   & 0.59  & 1.8   & 1.14  & 1.9             & 1.75           & 1.88            \\ 
S & 1.75  & 1.16  & 0.59  & 2.58  & N/A   & N/A             & 1.77           & 1.95            \\ \hline
\end{tabular}
\caption{Geometrical parameters and bond lengths of the considered interfaces A and S, all units are
in \AA.}
\label{tab-1}
\end{table}

\end{document}